\documentstyle[12pt,epsf,epsfig]{article}
\def\be{\begin{equation}}
\def\ee{\end{equation}}
\def\ba{\begin{eqnarray}}
\def\ea{\end{eqnarray}}
\def\ga{\mathrel{\raise.3ex\hbox{$>$\kern-.75em\lower1ex\hbox{$\sim$}}}}
\def\la{\mathrel{\raise.3ex\hbox{$<$\kern-.75em\lower1ex\hbox{$\sim$}}}}

\begin{document}

\begin{titlepage}
\pagestyle{empty}
\baselineskip=21pt
\rightline{TPI--MINN--00/37}
\rightline{astro-ph/0007079}
\rightline{July 2000}
\vskip.25in
\begin{center}

{\large{\bf Constraints over Cosmological Constant and Quintessence\\[1mm]
Fields in an Accelerating Universe}}
\end{center}
\begin{center}
\vskip 0.4in

{Vinod B. Johri*}
\vskip 0.2in
{\it
{Theoretical Physics Institute, School of Physics and Astronomy, \\
University of Minnesota, Minneapolis, MN 55455, USA}}
\vskip 0.5in
{\bf Abstract}
\end{center}
\baselineskip=18pt \noindent
A brief account of the current cosmological observations is given and their
implications for QCDM and $\Lambda$CDM cosmologies are discussed. The
nucleosynthesis and the galaxy formation constraints have been used to put
limits on $\Omega_\phi$ during cosmic evolution, and develop a realistic
approach to the tracking behaviour of quintessence fields. The astrophysical
constraints are applied to interpolate the value of the tracking parameter
$\epsilon \simeq 0.75$ at the present epoch and also
to find the lower and the upper limits for $\Lambda$
in the accelerating universe. It is shown that the transition from 
deceleration
to acceleration in the cosmic expansion occurs earlier in $\Lambda$CDM
cosmology compared to QCDM cosmology.
\\[3mm]
\vspace*{18mm}
\begin{flushleft}
\begin{tabular}{l} \\ \hline
{\small Emails: johri@hep.umn.edu}
\end{tabular}
\end{flushleft}

\end{titlepage}
\baselineskip=18pt

\section{Introduction}
The observational view of the universe has drastically changed during the
last ten years. Until a decade ago, the universe was supposed to be matter
dominated and the cosmic expansion was understood to be slowing down;
consequently, the Einstein de Sitter model was taken to be the standard
model of the observable universe. The latest cosmological observations
reveal a low mass density, spatially flat universe with accelerating
expansion. We shall briefly survey the current observations and apply
them to investigate the 'missing mass' problem in cosmology.

In cosmological theories, the Hubble Constant is one of the most important
observational parameters due to its sensitivity to the variations in the
cosmic energy density and the spatial curvature of the universe. According
to the current estimates \cite {1,2,3}, $H_o = 65\pm 10 \,km/Mpc/s $. As
regards the curvature of the universe, it was long predicted by the
inflationary scenarios that the observable universe must be spatially
flat. In the angular power spectrum of CMB \cite{4}, the location of the 
first
Doppler peak near $ l\simeq 200$ fortifies this view.
At the same time the precise measurements \cite{5,6,7,8} of the
density of matter derived from cluster abundances, baryon fraction
in clustered matter indicate a low mass density with
$\Omega_m\simeq 0.35$. It gives rise to the question as to how to account
for $65\% $ 'missing mass' in the universe, sometimes, referred to as 'dark
energy'.  The recent studies undertaken by  Supernova Cosmology  Project
Team \cite{9} and  High  Red-shift Search Team \cite{10} reveal that the
distant  SNe are fainter and thus more distant than expected for a 
decelerating
universe. It implies that the rate of cosmic expansion is accelerating
which 
in turn provides
empirical evidence of the existence of dark energy with negative
pressure. 
For
a spatially flat universe, the best values derived from the  analysis of 
combined
results of  \cite{9,10} are approximately $\Omega_m = 0.25\pm0.1$ and for 
dark
energy  $\Omega_X = 0.75\pm0.1(1 \sigma)$. These values are in excellent
agreement with  $\Omega_m$ derived from the baryon fraction in clustered
matter as discussed above. The most probable candidates for the dark
energy 
are
the cosmological
constant $\Lambda$ and the Quintessence fields -- the scalar fields
with evolving equation of state which, during the matter dominated phase,
acquire negative pressure and behave like $\Lambda_{eff}$.

As for the break-up of the material content of the universe, the
baryonic matter (BM) contributes only $\Omega_{BM} = 0.05$ as
determined from the precise measurement of deuterium abundance \cite{11,12}
in very distant hydrogen clouds. The neutrinos might contribute a small
fraction $\Omega_\nu \geq 0.003$. The remaining contribution
$\Omega_{DM}\approx 0.29$ comes from the dark matter (DM) which
determines the hierarchy
of the structure formation in the universe. The cold dark matter (CDM)
consists of particles like axions and neutralinos which move slowly and
cannot remove lumpiness on small scale; as such in CDM cosmology, the
structure formation follows bottom-up sequence i.e. the galaxies are
formed first followed by clusters and super-clusters. On the contrary,
the fast moving particles like neutrinos constitute hot dark matter
(HDM) which can remove lumpiness on small scale; as such HDM supports
top-down sequence of structure formation in the universe. The
astronomical observations made by the Keck 10-meter Telescope and
the Hubble Space Telescope reveal that most of the galaxies in the
universe formed between redshifts 2 to 4, that clusters formed at
redshift $z\leq 1$ and the superclusters are forming to-day. These
observations rule out HDM cosmology, restrict the contribution of
neutrinos to about $0.3\%$  as stated above and favour a CDM cosmology
for the observable universe.

It is important to distinguish between the dark matter and dark energy.
The dark matter may consist of exotic particles like axions and
neutralinos but it attracts and clumps like ordinary matter whereas
the dark energy does not clump; it repels matter. In the present paper,
we consider the role of two leading candidates of dark energy under the
sections QCDM cosmology and $\Lambda$CDM cosmology and discuss those
astrophysical/cosmological observations which constrain the magnitude
of the dark energy at the nucleosynthesis epoch, during galaxy formation
era and at the end of matter dominated (i.e. at the onset of acceleration)
era and their implications for the quintessence field and their tracking
behaviour. These constraints have also been used to put theoretical
limits on the magnitude of the cosmological constant $\Lambda$.

\section{ QCDM Cosmology}

Let us first consider CDM cosmology with Quintessence -- the rolling
scalar fields, with evolving equation of state, which acquire
repulsive character (owing to negative pressure) during the late evolution
of the universe. The quintessence in the present day
observable universe, behaves like $\Lambda_{eff}$ and may turn out to be 
the most
likely form of dark energy which induces acceleration in the cosmic
expansion.

Consider the homogeneous scalar field $\phi(t)$ which interacts with matter
only through gravity. The energy density $\rho_\phi$ and the pressure 
$p_\phi$
of the field are given by
\be
\rho_\phi\, = \frac 12\,\dot\phi^2 + V(\phi)
\ee
\be
p_\phi\, = \frac 12\,\dot\phi^2 - V(\phi)
\ee
The equation of motion of the scalar field
\be
\ddot{\phi} + 3H\dot{\phi} + V'(\phi)\, = 0, \qquad
V'(\phi)\equiv\frac{dV}{d\phi}
\ee
leads to the energy conservation equation
\be
\dot{\rho}_\phi + 3H(1+w_\phi)\rho_\phi\, = 0
\ee
where $w_\phi\equiv\frac{p_\phi}{\rho_\phi}$ and
$H\equiv\frac{\dot{a}}{a}$ 
is
the
Hubble constant. Accordingly, $\rho_\phi$ scales down as
\be
\rho_\phi\,\sim {a}^{-3(1+w_\phi)}  ,  \quad -1\leq w_\phi \,<\frac 13
\ee
Obviously, the scaling of $\rho_\phi$ gets slower as the potential energy
$V(\phi)$ starts dominating over the kinetic energy $\frac 12 \dot{\phi}^2$
of the scalar field and $w_{\phi}$ turns negative.

Since there is minimal interaction of the scalar field with matter and 
radiation,
It follows from Eq.(4) that the energy of matter  and radiation is
conserved 
separately as
\be
\dot\rho_n + 3H(1 + w_n)\rho_n = 0
\ee
Accordingly
\be
\rho_n\,\sim a^{-3(1+w_n)}
\ee
where $\rho_n$ is the energy density of the dominant constituent (matter or
radiation) in the universe with the equation of state $p_n = w_n\,\rho_n$
where $w_n = \frac 13$ for radiation and $w_n = 0$ for matter.

Although, the scalar field is non-interactive with matter, it affects the
dynamics of cosmic expansion through the Einstein field equations. Assuming
large scale spatial homogeneity and isotropy of the universe, the field
equations for a flat Friedmann model are
\be
      H^2 \, = \frac{\rho_n + \rho_\phi}{3M_p^2}
\ee
and
\be
   \frac{2\ddot {a}}{a}\, =  -\frac{\rho_n+\rho_\phi +3p_n +3p_\phi}{3M_p^2}
\ee
where $M_p =2.4\times 10^{18}$ GeV is the reduced Planck mass.

Denoting the fractional density of the scalar field by $\Omega_\phi\equiv
\frac{\rho_\phi}{\rho_n+\rho_\phi}$ and that of the matter/radiation
field 
by
$\Omega_n\equiv\frac{\rho_n}{\rho_n+\rho_\phi}$, equations (8) and (9)
may 
be
rewritten as
\be
      \Omega_n + \Omega_\phi\, = \, 1
\ee
and
\be
2\frac{\ddot a}{a} = -\frac{\rho_n}{3M_p^2}\,[(1+3w_n)+(1+3w_\phi)
\frac{\Omega_\phi}{\Omega_n}]
\ee

The relative growth of $\Omega_\phi$ versus $\Omega_n$ during the cosmic
evolution is given by
\be
\frac{\Omega_\phi}{\Omega_n} = \frac{\Omega_\phi^0}{\Omega_n^0}\,
\biggl(\frac{a}{a_0}\biggr)^{3\epsilon}
\ee
where the tracking parameter $\epsilon\equiv w_n - w_\phi$ and
$\Omega_\phi^0$, $\Omega_n^0$  denote the values of $\Omega_\phi$ and 
$\Omega_n$
at the present epoch $ (a = a_0)$. As discussed in section 1, the  $SNe I_a$
observations suggest that $\Omega_\phi^0
\simeq 2\Omega_n^0$; consequently Eq.(12) may be expressed in terms of the
red-shift $z$ as below
\be
2(\Omega_\phi^{-1} - 1)  =\, (1+z)^{3\epsilon}
\ee

If we insist that the scalar field, regardless of its initial value,  should
behave like $\Lambda_{eff}$ today, it must obey tracking conditions
\cite{13,14,15}
with wide ramifications for quintessence fields already discussed in detail
\cite{16}-\cite{21}. In nutshell, tracking
consists in synchronised scaling of $\rho_\phi$ and $\rho_n$ along a common
evolutionary track so as to ensure the restricted growth of $\Omega_\phi$
during the cosmic evolution in accordance with the observational (both
astrophysical and cosmological) constraints. As discussed in \cite{15}, the
tracking
parameter $\epsilon$ plays a vital role in monitoring the desired growth of
$\Omega_\phi$. The existence theorem for tracker fields \cite{15} requires
$\epsilon$ to satisfy the condition
$\frac{\epsilon\Omega_n}{2(1+w_\phi)} < 
1$
and also to conform to cosmological constraints mentioned therein. Here we
investigate the consequent limits on the tracking parameter $\epsilon$ 
imposed
by these constraints to ensure tracking by a quintessence field.

According to the Eq. (13),  $\Omega_\phi$ depends both upon the redshift $z$
and the tracking parameter $\epsilon$. From a general functional form
$\Omega_\phi = f(z, \epsilon)$, it is difficult to chart out
$\Omega_\phi -
z$ variation through tracking unless we are able to fix the value of
$\epsilon$ corresponding to known $\Omega_\phi$ at certain points in the 
phase
space of $z - \epsilon$. This is achieved with the help of the astrophysical
constraints discussed in this paper. Having evaluated $\Omega_\phi$ at some
typical points $(z_0, \epsilon_0)$, we can interpolate
$\Delta\Omega_\phi$ 
at
neighbouring points $(z_0 + \Delta z_0 , \epsilon_0 +
\Delta\epsilon_0)$. 
Using
this technique, realistic tracking diagrams of $\Omega_\phi - z$ variation,
$w_\phi - z$  variation and
$\epsilon - z$ variation may be drawn as shown in the figures 1,2 and 3.

\begin{figure}
\begin{center}
\mbox{\epsfig{file=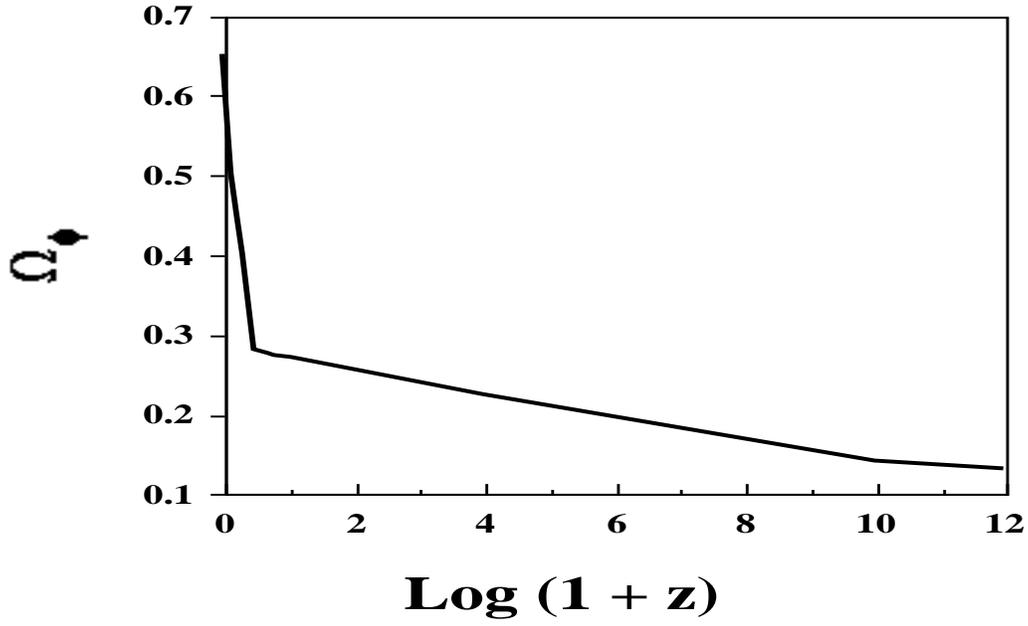,height=8.3cm,width=14cm}}
\end{center}
\caption[.]{\label{fig:bsg1}\it
Variation of $\Omega_\phi$ versus Redshift $z$ in QCDM Cosmology
assuming $H_o=65$ Km/Mpc/s.}
\end{figure}

\begin{figure}
\begin{center}
\mbox{\epsfig{file=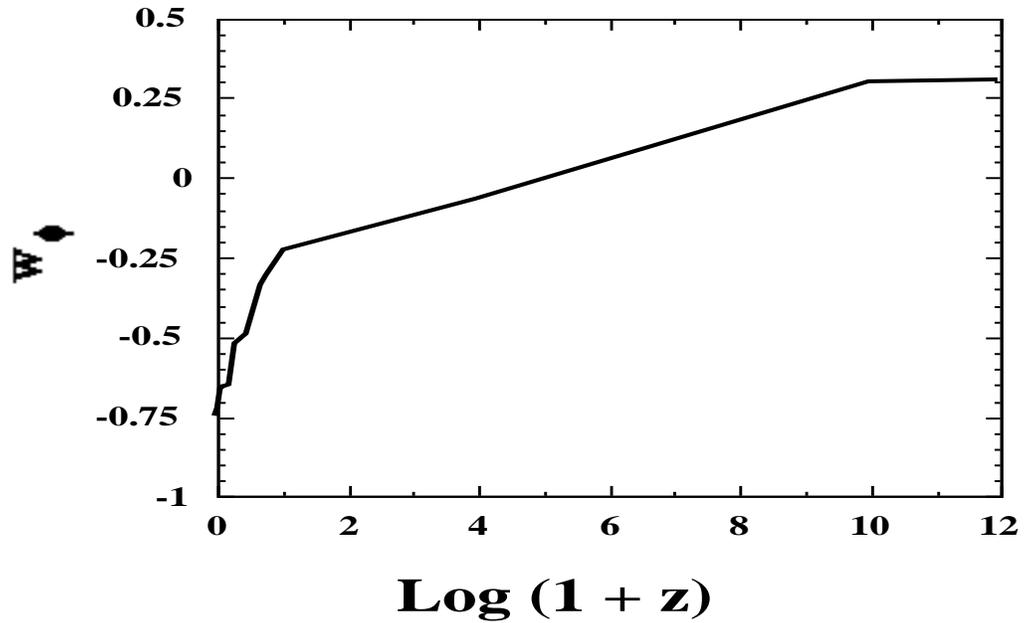,height=8.3cm,width=14cm}}
\end{center}
\caption[.]{\label{fig:bsg2}\it
Evolution of the equation of state of the quintessence field in QCDM 
Cosmology
($H_o=65$ Km/Mpc/s).}
\end{figure}

\begin{figure}
\begin{center}
\mbox{\epsfig{file=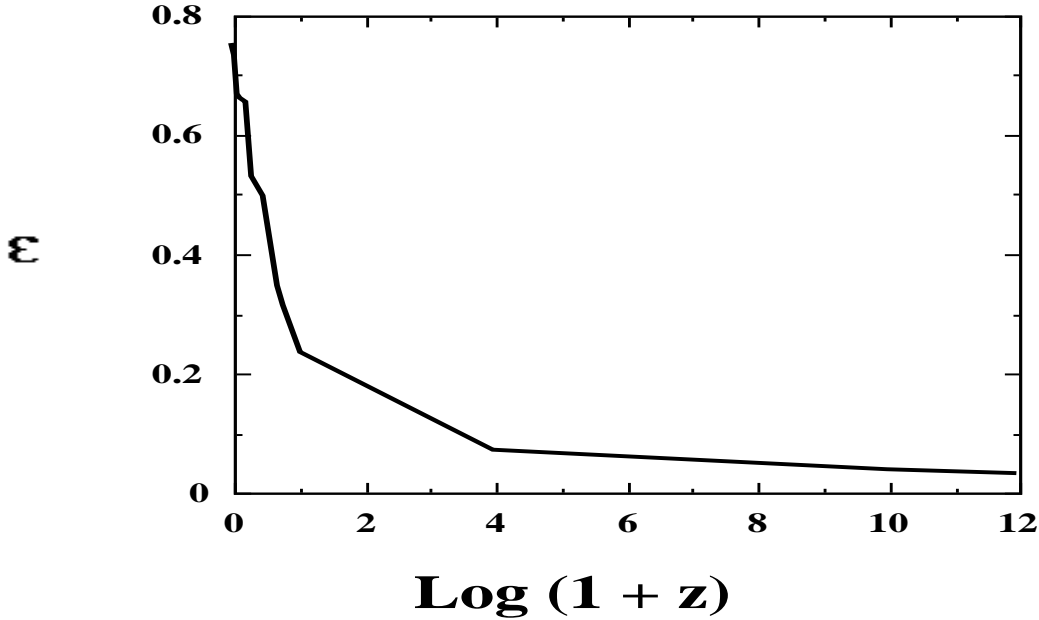,height=8.3cm,width=14cm}}
\end{center}
\caption[.]{\label{fig:bsg3}\it
Variation of the tracking parameter $\epsilon$ versus Redshift $z$
in QCDM Cosmology ($H_o=65$ Km/Mpc/s).}
\end{figure}

In this connection, the following differential relation, derived from
Eq. 
(13)

\be
  -  \frac{\Delta\Omega_\phi}{3\epsilon\Omega_n\Omega_\phi} \, =
\frac{\Delta z}{1+z}
  + \frac{\Delta\epsilon}{\epsilon} \ln(1+z).
\ee
is  found quite useful in interpolating the increment
$\Delta\Omega_\phi$ in
terms of the increments $\Delta z$ and $\Delta\epsilon$. It is noteworthy
that the contribution of the term $\frac{\Delta z}{1+z}$ is very small
compared to the contribution of $\frac{\Delta\epsilon}{\epsilon}$ at high
  redshifts.

  Let us now reconsider the astrophysical constraints discussed in our 
previous
paper \cite{15}, try to refine them and examine their implications for
quintessence
fields.

{\bf 1. The Nucleosynthesis Constraint.}
   The first constraint on $\Omega_\phi$ during the cosmic evolution
comes 
from
the
   helium abundance at the nucleosynthesis epoch $( z\sim 10^{10})$. The
presence
   of an additional component of energy in the form of quintessence
field 
with
   energy density $\rho_\phi$ results in an increase in the value of the 
Hubble
   constant $H$ as given by the  differential of the Friedmann equation
   \be
   \frac{2\delta H}{H} \, = \frac{\delta\rho}{\rho}\, = 
\frac{\rho_\phi}{\rho}.
   \ee
   This, in turn, yields a higher ratio of neutrons to protons at the 
freeze-out
   temperature (1 MeV) of the weak interactions and a consequent higher
percentage
   of the helium abundance in the universe. Assuming the existence of
three 
known
   species of neutrinos, the nucleosynthesis calculation \cite{22} yields
   \be
   \frac{\delta\rho}{\rho} \, = \, \frac{7(N_\nu - 3)}{43}
   \ee

\begin{figure}
\begin{center}
\mbox{\epsfig{file=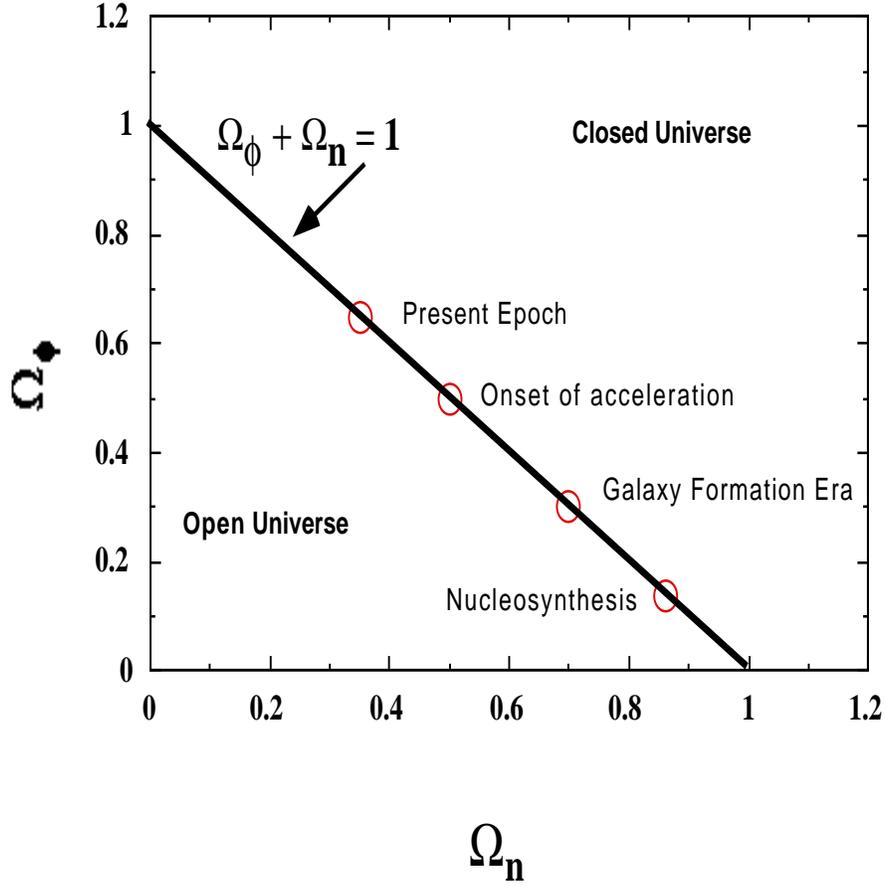,height=12cm,width=12cm}}
\end{center}
\caption[.]{\label{fig:bsg4}\it
The important cosmic events corresponding to the astrophysical constraints
are marked by circles on the thick line (representing spatially flat 
universe)
in QCDM Cosmology.}
\end{figure}

   Since the number of neutrino species $N_\nu < 4$, we arrive at the 
constraint
   $\frac{\delta\rho}{\rho} < \frac{7}{43}$. If the contribution 
$\delta\rho$
   comes from the quintessence field instead, the above constraint 
translates
into
   nucleosynthesis constraint on $\Omega_\phi$ as follows
   \be
   \Omega_\phi = \frac{\rho_\phi}{\rho_n + \rho_\phi} < \frac{7}{50} = 0.14
   \ee
   The corresponding value of the tracking parameter is $\epsilon\leq0.035$.

{\bf 2. Galaxy Formation Constraint.}
     According to the current estimates in CDM cosmology, the galactic 
structure
     formation takes place   between the redshift $z = 4$ and $z = 2$. The
clumping
     of matter into galaxies demands the dominance of gravitational 
attraction
     during this period. Therefore, the repulsive force of quintessence
must 
be
     relatively weak and $\Omega_\phi$ must be reasonably less than 0.5 
during
     the galaxy formation era. Interpolation from Eqs. (13) and (14)
shows 
that
     $0.33\leq \epsilon \leq 0.5$ during the galaxy formation era.

{\bf 3. Present Epoch.}
    Two recent surveys \cite{9,10} based on $SN_e  I_a$ measurements
    predict accelerating cosmic expansion  with $\Omega_\phi\simeq 0.65\
    \pm 0.05$ at the present epoch (z=0). The constraint $\ddot a >0 $ 
inserted
    in Eq. (11) leads to
    $\epsilon > 0.5$ at the present epoch. Interpolation with
    the help of Eq.(14) places $\epsilon$ around 0.75.

{\bf 4. Transition to Accelerated Expansion (Quintessence Dominated Era).}
     The onset of acceleration( $\ddot a \ga 0$) in the observable universe
takes place around
     the value of $\Omega_\phi\geq 0.5$ which corresponds to
     $\epsilon\sim0.66$
     from Eq.(11) at a redshift of $z= 0.419$ from Eq.(13).

In figure 4, the cosmic events are depicted sequentially by the thick
line 
in
the $\Omega_n - \Omega_\phi$ diagram. The events corresponding to the above
constraints
are marked by circles.

\section{ $\Lambda CDM$ Cosmology}

In this section , we regard the universe to be filled up with a mixture
of 
cold
dark matter and the vacuum energy represented by the cosmological constant
  $\Lambda$ which plays the role of dark energy and tends to accelerate the
  expansion of the universe.

  In the presence of $\Lambda$ term, the Einstein field equations for a
spatially
  flat Friedmann universe are given by
  \be
       H^2 = \frac{8\pi G}{3} (\rho_n + \rho_\Lambda)
  \ee
  and
  \be
  \frac{2\ddot a}{a} = - \frac{8\pi G}{3} (\rho_n + 3p_n/c^2 - 
2\rho_\Lambda)
  \ee
  where $\rho_\Lambda = \frac{\Lambda c^2}{8\pi G}$ and the Newtonian
Gravitational
  constant $G=6.6\times 10^{-8}$ cgs units.
  In case of matter dominated universe $(p_n \sim 0)$, the Eqs. (18) and 
(19)
  may be rewritten as
  \be
       \Omega_n  +  \Omega_\Lambda  =  1
  \ee
  and
  \be
       q  =  \frac 12 \Omega_n  -  \Omega_\Lambda
  \ee
  where $q \equiv  - \frac{\textstyle \ddot a}{\textstyle a H^2}$ is the
  deceleration parameter and
  $\Omega_\Lambda \equiv \frac{\textstyle \rho_\Lambda}{\textstyle
\rho_\Lambda+\rho_n}
  =\frac{\textstyle \Lambda c^2}{\textstyle 3 H^2}$.

  It is obvious from Eq. (21) that the cosmic expansion slows down as
long as
  $\Omega_n >  2 \Omega_\Lambda$ (i.e $\rho_n > 2\rho_\Lambda$); the 
clumping
  of matter into galaxies takes place during this period. Transition to
  accelerated expansion occurs when the deceleration parameter $q\leq 0$ 
which
  corresponds to $\Omega_\Lambda \geq \frac 13$. It marks the beginning
of 
the
  $\Lambda$-dominated phase during which the universe goes on expanding 
faster
  and faster and ultimately enters the de Sitter phase of exponential 
expansion.
  This is borne out clearly from the analytical solution of the field 
equations
  (18) and (19) during matter dominated phase
  \be
           a \sim \,  \sinh^{2/3}(\frac 32 \sqrt{\Lambda/3}\,ct)
   \ee
    It is noteworthy that this solution reduces to Einstein de Sitter 
solution
    $ a\sim t^{2/3}$ in the limit $\Lambda\rightarrow 0$ and goes back
to de
Sitter
    expansion as $\rho_n \rightarrow 0$.

  It leads to
  \be
  H  =  c\sqrt{\Lambda/3} \coth(\frac 32\,\sqrt{\Lambda/3}\,ct)
  \ee
  and
  \be
        q = \frac 13 - \tanh^2 (\frac 32\,\sqrt{\Lambda/3}\,ct)
  \ee

  The astrophysical constraints discussed in the previous section apply to
  $\Lambda$ as well. Since $\Lambda$ remains constant throughout, these
constraints
  put theoretical limits on the plausible values of $\Lambda$. For 
instance, the
  clumping of matter into galaxies can take place during the cosmic 
deceleration
   phase ($q > 0$) when the gravitational attraction is dominant over cosmic
   repulsion. It follows from Eq. (24) that during the galaxy formation era
   \be
   \tanh^2\,\biggl(\frac 32\,\sqrt{\Lambda/3}\,ct \biggr)\, < \, \frac 13
   \ee
   Expressed in terms of redshift $z$ with age $t_0$ of the universe
taken 
as
   13 billion years, we get
   \be
   \Lambda  < 4.2\times 10^{-57} (1+z)^3
   \ee
   Assuming that the galaxy formation continues up to redshift $z = 2$, we
   derive the upper limit of $\Lambda$
   \be
   \Lambda  <  33.5\times 10^{-57}
   \ee

   The transition to accelerated expansion takes place at $ t = t_c $
corresponding
   to redshift $z = z_c $ as given by Eq. (24) with $ q = 0 $.
   At the present epoch $(z = 0)$, the cosmic expansion is accelerating
   ($q < 0$).
   This yields the lower limit of $\Lambda$ as given by
   \be
   \Lambda  >  4.2\times 10^{-57}
   \ee

   According to the constraint 4 in Section 2, it was found that the
   transition to accelerated expansion (scalar field dominated) phase in 
QCDM
cosmology
   occurs at
   $ z_c \sim 0.419 $. Since the transition in $\Lambda$CDM cosmology takes
place earlier,
   we can take $z_c\simeq 0.5$. This corresponds to material density
$\rho_n 
=
7.415\times
   10^{-30}$ gm/cm$^3$. At the point of onset of acceleration ($q = 0$),
we 
have
   by Eq. (21)
   \be
   2\rho_\Lambda  =  \rho_n  =  8.43\times 10^{-30} gm/cm^3
   \ee
   It yields the value
   \be
   \Lambda  =  6.99\times 10^{-57}
   \ee
   which is in good agreement with the observational estimate $\Lambda  =
   \frac{\textstyle 3 H_0^2 \Omega_\Lambda}{\textstyle c^2}  =
   7.74\times 10^{-57}$
   based on the value of the Hubble constant $H_0 = 65$ km/Mpc/s and
   $\Omega_\Lambda  =  0.65$.

\section{Conclusions}

The astrophysical constraints on $\Omega_\phi$, discussed in Section 2,
enable us to present a realistic picture of the scaling of the quintessence
energy during tracking and plot the variation of $\Omega_\phi$ versus
$\epsilon ,\,w_\phi$ and z (redshift). Whereas both QCDM
cosmology and $\Lambda$CDM cosmology are fairly compatible with the
recent CMB observations and power spectrum of galaxy clustering, QCDM
has the advantage in fitting constraints from high redshift supernovae,
gravitational lensing and structure formation at large redshifts
\cite{17}. Another distinguishing feature of QCDM cosmology is that the
onset of cosmic acceleration takes place when $\Omega_\phi = 0.5 $
whereas in $\Lambda$CDM cosmology, $\Omega_\Lambda = \frac13$ marks the
onset of acceleration. Taking the observational value
$\Lambda = 7.74\times 10 ^ {-57}$, the transition to the accelerating
expansion phase in $\Lambda$CDM cosmology occurs when
$\rho_m = 9.3\times 10^ {-30} gm/cm ^3 $ which corresponds to
$z_c \sim 0.54$. In QCDM cosmology, as discussed in Section 2,
the transition occurs at $z_c = 0.419$. It means that the transition
to accelerating expansion occurs earlier in $\Lambda$CDM cosmology
than in QCDM cosmology. Again, in $\Lambda$CDM cosmology, the universe
ends up in inflationary phase with exponential expansion whereas in
QCDM cosmology, the ultimate fate of the universe is inflation with
exponential or hyperbolic expansion, depending upon the form
of the scalar potential.

{\large\bf Acknowledgments}
This work was supported in part by UGC grant from India. The author
acknowledges useful
discussions and valuable help of Keith Olive and Panagiota Kanti and
hospitality of
Theoretical Physics Institute, University of Minnesota.

* Permanent Address: Department of Mathematics and Astronomy, Lucknow
University,
   Lucknow 226007. India.

\end{document}